\documentclass[preprint,12pt]{elsarticle}

\usepackage[utf8]{inputenc}
\usepackage[T1]{fontenc}
\usepackage{lmodern}
\usepackage[english]{babel}
\usepackage{csquotes}
\usepackage{amsmath,amssymb}
\usepackage{graphicx}
\usepackage{booktabs}
\usepackage{tabularx}
\usepackage{multirow}
\usepackage{float}
\usepackage{xcolor}
\usepackage{microtype}
\usepackage{xurl}
\usepackage[colorlinks=true,linkcolor=blue!60!black,citecolor=blue!60!black,urlcolor=blue!60!black]{hyperref}

\graphicspath{{figures/}}

\newcommand{\Cos}{Cosmos~2612}
\newcommand{\Ice}{ICEYE-X36}

\journal{Acta Astronautica}

\begin{document}

\begin{frontmatter}

\title{Analysis of the Cosmos~2612 -- ICEYE-X36 rendezvous and proximity operation from public orbital element data}

\author{Brayan Nicolas Buitrago Montenegro\corref{cor1}}
\ead{brbuitragom@unal.edu.co}
\cortext[cor1]{Corresponding author. ORCID: \href{https://orcid.org/0009-0004-8742-171X}{0009-0004-8742-171X}}

\affiliation{organization={Observatorio Astronómico Nacional, Facultad de Ciencias, Universidad Nacional de Colombia},
            addressline={Av. Carrera 30 No. 45-03, Ciudad Universitaria},
            city={Bogotá D.C.},
            postcode={111321},
            country={Colombia}}

\begin{abstract}
We analyse the October 2026 close approach between \Cos{} (NORAD~68762) and the commercial radar satellite \Ice{} (NORAD~59103), both in sun-synchronous low Earth orbit, to determine whether it was a rendezvous and proximity operation (RPO) or the result of natural orbital evolution. The public general perturbations (GP) element history from April to October 2026 was propagated with SGP4, and the prediction error was measured empirically for each object. The null hypothesis of natural evolution, tested with manoeuvre-free propagations, is rejected for all 18 parameter combinations: the encounter required manoeuvres by \Cos{}, whereas those of \Ice{} are not needed to reproduce it. Closest approach occurred on 18 October at 06:33~UTC, below 0.79~km (95th percentile), with a local relative velocity of $\approx$0.08~m/s and 41.4~h continuously within 5~km. It was the final stage of a sequence that began in May with a plane match, followed by a phasing drift, the approach and a manoeuvred retreat, with a minimum cumulative $\Delta v$ of 174--180~m/s. The data cannot resolve the minimum distance below their accuracy, nor establish whether the operation was cooperative or its purpose.
\end{abstract}


\end{frontmatter}

\section{Introduction}

Rendezvous and proximity operations (RPO) are manoeuvres by which a spacecraft approaches another object in orbit in a planned way and remains in its vicinity for a given time. They have civil and commercial uses, such as inspection, servicing, refuelling or life extension of satellites, and also military and intelligence uses, such as surveillance or co-orbital anti-satellite weapons \cite{samson2026}.

RPO have already been documented and classified according to the observed behaviour. One of the proposed categories is \emph{provocative behaviour}: an RPO carried out adversarially, because of the physical threat it poses to another satellite or for information gathering. Without coordination between operators, this kind of activity also poses a collision risk to the target satellite and to nearby satellites \cite{kelecy2022}. RPO for military or intelligence purposes are not exclusive to one country: they have been documented, for example, in programmes of the United States \cite{samson2026} and in Russian activities in low Earth orbit (LEO) and geostationary orbit (GEO) \cite{mowthorpe2026}.

Public catalogues of orbital elements make it possible, with known limitations, to reconstruct these events independently. This possibility, however, is not the same for all objects. The Space-Track data provider warns that the public catalogue may not include orbital elements for all objects, among other reasons for national security, although all objects are tracked at least once a day \cite{spacetrackfaq}. Consequently, RPO involving military or intelligence satellites whose elements are not published cannot be reconstructed with public data, and the open record of these events depends on which objects appear in the catalogue. In the case studied here, the elements of both objects are public.

Cooperative RPO have a long track record: Barnhart et al. reviewed the rendezvous missions carried out up to 2020 to propose safety metrics applicable to future on-orbit services \cite{barnhart2020}, and the design of services such as active debris removal includes the phases of an RPO with a non-cooperative target, from far-range rendezvous to inspection \cite{borelli2023}. Undeclared RPO have been reconstructed from public data. Serrano et al. analysed several cases in GEO and LEO with TLEs, including Kosmos~2542 and USA~245 (2019--2020) and Kosmos~2558 and USA~326 (2022), in which the Russian satellite occupied an orbital plane almost identical to that of the target and made periodic approaches at distances of tens of kilometres; they also note that, with about one TLE per day, the exact manoeuvres cannot be determined \cite{serrano2024}. Du et al. systematically searched the public TLE catalogue for 2020--2025 for RPO candidates, combining orbital-plane similarity, distance and persistence of proximity, and identified a small number of unexplained cases between satellites of different countries \cite{du2026}. These works describe the trajectories or apply fixed distance thresholds, but they do not pose a formal test between natural evolution and an induced approach, with a prediction error measured for each object, nor do they determine which of the two objects performed the necessary manoeuvres. That is the approach of the present work.

\Ice{} is a synthetic aperture radar (SAR) Earth-observation satellite owned by ICEYE~US, in sun-synchronous orbit, launched on 4 March 2024 on a Falcon~9 rocket from Vandenberg Space Force Base, California \cite{isstracker59103}. On 19 January 2026, ICEYE and a customer from the Ministry of Defence of Ukraine signed an agreement that expands that ministry's access to high-resolution SAR imagery from the company's constellation \cite{iceye2026}; the press release does not identify which satellites provide the service. According to Scientific American, \Ice{} has been used by the Ukrainian military \cite{randolph2026}. \Cos{} is a Russian military satellite launched on 16 April 2026 at 23:17~UTC (17 April, Moscow time) on a Soyuz rocket with a Volga upper stage, in a launch from which six payloads were identified (Cosmos~2609 to 2614); the correspondence between \Cos{} and the object NORAD~68762 (international designator 2026-083H) is considered presumed \cite{zak2026}.

After the October 2026 encounter, several media outlets reported that \Cos{} had approached \Ice{}. Scientific American, based on public tracking data, describes the separation as at most a few hundred metres, with estimates down to 30~m, and calls the event the closest known deliberate and unsolicited interaction between two active spacecraft in low orbit \cite{randolph2026}. According to a LeoLabs analysis cited by CNN, the separation reached $\approx$500~m during part of the event, with an uncertainty due to the repeated manoeuvres and the tracking geometry \cite{cnn2026}. The Washington Times mentions an estimated distance of $\approx$26~m \cite{washingtontimes2026}. AeroTime states that the approach may have been a few hundred metres or less, at around 01:35~UTC, and that the uncertainty of public orbital data prevents a reliable calculation of the minimum separation \cite{aerotime2026}. These claims motivate the present study but are not adopted as a premise: the analysis is set up as a hypothesis test whose result could, in principle, have been that the encounter was natural.

The aim of this work is to answer three questions with public data and reproducible methods: (i)~can the approach be explained by the natural evolution of the orbits, or did it require manoeuvres?; (ii)~if it did, by which object?; and (iii)~what were the geometric and kinematic characteristics of the encounter, and which of them cannot be determined with the accuracy of the data?

\section{Data and methods}

\subsection{Data}

We used the general perturbations (GP) orbital element history from Space-Track \cite{spacetrack}, distributed in the TLE (\emph{two-line element set}) format, for both objects between 1 April and 28 October 2026 (Table~\ref{tab:datos}). When several element sets shared the same epoch, the most recently created one was kept (54 duplicates removed from 1408 records). The analysis window of the encounter is 1--27 October 2026; the period before 1 September is used as the \emph{reference period} to calibrate the propagation error and the manoeuvre detector.

\begin{table}[H]
\centering
\caption{GP data used.}
\label{tab:datos}
\resizebox{\textwidth}{!}{%
\begin{tabular}{lcccc}
\toprule
Object & NORAD & No. of TLEs & First / last epoch (2026) & TLEs per day \\
\midrule
\Ice{}  & 59103 & 758 & 01-04 / 27-10 & 3.62 \\
\Cos{} & 68762 & 596 & 18-04 / 27-10 & 3.10 \\
\bottomrule
\end{tabular}}
\end{table}

\subsection{Propagation and empirical SGP4 error}

Positions and velocities were obtained with the SGP4 model (\emph{Simplified General Perturbations~4}) using the \texttt{sgp4} library \cite{rhodes_sgp4}, in the TEME frame (\emph{True Equator, Mean Equinox}), sampled every 60~s. To reconstruct the observed trajectory, at each instant the most recent TLE with an earlier epoch is used (causal selection); the manoeuvre-free propagations of the hypothesis test (Section~\ref{sec:contraste}) instead use a fixed TLE that is not updated. The mean elements are obtained from each TLE: semi-major axis $a$, perigee and apogee altitudes $h_{p,a} = a(1\mp e) - R_\oplus$, where $R_\oplus$ is the Earth's equatorial radius, inclination $i$, eccentricity $e$ and the atmospheric drag term $B^*$.

TLEs do not include a formal estimate of their uncertainty (covariance), so the SGP4 prediction error was measured empirically for each object. In the reference period, each TLE was propagated to the epoch of a later TLE of the same object located $L \pm 0.25$~days ahead, and the distance between the propagated position and the one given by the later TLE at its own epoch was computed. Because the later TLE incorporates more recent observations, this distance approximates the error made when predicting the position $L$ days ahead. This comparison between successive TLEs, known as pairwise differencing, is an established procedure for estimating TLE errors when no covariance is available, and weighted variants based on the same principle also exist \cite{geul2017}. Pairs between which the detector identified a manoeuvre (Section~\ref{sec:maniobras}) were excluded, because their difference reflects the orbit change rather than the model error. The 95th percentile of the distribution obtained for each $L$, $\varepsilon_{95}(L)$, is used as the reference uncertainty (Table~\ref{tab:error}, Figure~\ref{fig:error}). For the relative position between the two objects, whose errors are assumed independent, the individual errors are combined in quadrature, $\varepsilon_{95}^{\mathrm{rel}}(L) = \sqrt{\varepsilon_{95,\mathrm{ICEYE}}(L)^2 + \varepsilon_{95,\mathrm{Cosmos}}(L)^2}$ (row \emph{Relative} of Table~\ref{tab:error}).

\begin{table}[H]
\centering
\caption{SGP4 prediction error (95th percentile, km) as a function of propagation time $L$, measured in the manoeuvre-free reference period.}
\label{tab:error}
\begin{tabular}{lccccccccc}
\toprule
$L$ (days) & 0.25 & 0.5 & 1 & 2 & 3 & 5 & 7 & 10 & 14 \\
\midrule
\Ice{}  & 0.31 & 0.47 & 1.04 & 2.65 & 5.49 & 16.0 & 32.1 & 73.6 & 175 \\
\Cos{} & 0.32 & 0.50 & 0.76 & 1.52 & 2.93 & 8.33 & 21.2 & 47.5 & 120 \\
Relative & 0.44 & 0.69 & 1.28 & 3.05 & 6.23 & 18.0 & 38.4 & 87.6 & 213 \\
\bottomrule
\end{tabular}
\end{table}

\subsection{Manoeuvre detection and velocity-change estimation}\label{sec:maniobras}

An orbital manoeuvre is a firing of the propulsion system that changes the satellite's velocity. Its magnitude is expressed as the velocity change $\Delta v$ it produces, which is the usual measure of the cost of a manoeuvre because it determines the propellant consumed. Since manoeuvres are not published, they must be inferred from the changes that appear between successive TLEs.

For each pair of consecutive TLEs we computed the change in semi-major axis $\Delta a$, the change in inclination $\Delta i$ and the propagation residual $\rho = \lVert \mathbf r_{k-1}(t_k) - \mathbf r_k(t_k)\rVert$ normalised by the interval between epochs. Each series was transformed into a robust score, which replaces the mean and standard deviation with the median and the median absolute deviation (MAD), which are less sensitive to outliers \cite{rousseeuw2018}:
\begin{equation}
z = \frac{x - \operatorname{median}(x_{\mathrm{ref}})}{1.4826\,\operatorname{MAD}(x_{\mathrm{ref}})},
\end{equation}
where MAD is the median of the absolute deviations from the median and the factor 1.4826 makes it consistent with the standard deviation for a Gaussian distribution \cite{rousseeuw2018}. As in other TLE-based manoeuvre detectors, the threshold is derived from the statistics of each object \cite{cui2024}; here it is computed only from its reference period. Pairs with $|z| > 5$ were flagged, a more conservative threshold than the indicative value of 2.5 \cite{rousseeuw2018}, to limit false alarms due to TLE noise.

Flagged pairs less than 1.5~days apart are grouped into one event. The interval between consecutive TLE epochs has a median of 6.4~h for \Ice{} and 8.0~h for \Cos{}, and a 99th percentile of 25.5 and 23.9~h, respectively. A threshold of 1.5~days (36~h) exceeds that percentile, so two flagged pairs separated by a single normal interval between TLEs, as happens when a manoeuvre or a multi-burn transfer is spread over consecutive pairs, are grouped into the same event, while manoeuvres several days apart remain separate events. The grouping does not enter the hypothesis-test rules directly (Section~\ref{sec:contraste}); it affects the manoeuvre table and the exclusion of pairs in the error calculation. With thresholds between 0.5 and 3~days, the cumulative $\Delta v$ of \Cos{} since May varies between 179.6 and 179.7~m/s (gross) and between 174.0 and 176.9~m/s (net).

$\Delta v$ is estimated with an impulsive, near-circular approximation,
\begin{equation}
\Delta v_a \approx \frac{v}{2a}\,|\Delta a|, \qquad \Delta v_i = 2v\sin\frac{|\Delta i|}{2}, \qquad v = \sqrt{\mu/a},
\end{equation}
where $\mu$ is the Earth's gravitational parameter, $\Delta v_a$ is the impulse associated with the change in semi-major axis and $\Delta v_i$ the one associated with the change in inclination; the expression for $\Delta v_a$ follows from the linear relation between the change in semi-major axis and a tangential impulse \cite{cui2024}. Since both components are perpendicular (along-track and normal to the orbital plane), they are combined in quadrature, $\Delta v = \sqrt{\Delta v_a^2 + \Delta v_i^2}$, which is equivalent to a single burn producing both changes. $\Delta v$ is reported in two ways: \emph{net}, between the first and last TLE of the event, and \emph{gross}, as the sum of $\Delta v$ over each pair of consecutive TLEs, which prevents burns of opposite sign from cancelling out. Both are lower bounds, because several manoeuvres between two TLEs are observed as one. The detector was applied identically to both objects.

\subsection{Relative motion}

Let $\mathbf r_A, \mathbf v_A$ be the position and velocity of \Ice{} and $\mathbf r_B, \mathbf v_B$ those of \Cos{}. The relative position is $\Delta\mathbf r(t) = \mathbf r_B(t) - \mathbf r_A(t)$ and the distance is $d(t) = \lVert\Delta\mathbf r(t)\rVert$, computed every 60~s. The TCA (\emph{time of closest approach}) is the instant of closest approach, i.e. the minimum of $d(t)$ in the analysis window, refined to a resolution of 1~s.

The separation is described in the local RIC frame of \Ice{}, whose axes are the radial direction, the along-track (\emph{in-track}) direction and the normal to the orbital plane (\emph{cross-track}); this frame is also called RTN or the Hill frame: $\hat{\mathbf R} = \mathbf r_A/r_A$, $\hat{\mathbf C} = \mathbf h_A/h_A$, $\hat{\mathbf I} = \hat{\mathbf C}\times\hat{\mathbf R}$, with $\mathbf h_A = \mathbf r_A \times \mathbf v_A$ \cite{seong2013}. The difference of inertial velocities includes the rotation of the frame ($\approx n\,d$, with $n$ the mean motion, for two objects on the same orbit separated by a distance $d$), so the relative velocity is reported in the local frame, which rotates with $\boldsymbol\omega = \mathbf h_A/r_A^2$:
\begin{equation}
\mathbf v_{\mathrm{loc}} = (\mathbf v_B - \mathbf v_A) - \boldsymbol\omega \times (\mathbf r_B - \mathbf r_A).
\end{equation}
This quantity is representative at separations of a few kilometres; at hundreds of kilometres it retains contributions from the eccentricity and from the orbital curvature, which the rectilinear frame does not capture.

\subsection{Hypothesis test}\label{sec:contraste}

Two hypotheses were defined:
\begin{itemize}
\item \textbf{H0 (natural evolution):} the relative trajectory around the TCA is the ballistic, manoeuvre-free continuation of the prior orbital state of both objects.
\item \textbf{H1 (induced approach):} the observed relative trajectory cannot be obtained without manoeuvres, and the discrepancy is explained by manoeuvres of at least one of the objects.
\end{itemize}

\paragraph{R1, need for manoeuvres.} For each lead time $L \in \{0.5,\,1,\,2,\,3,\,5,\,7,\,10\}$~days (\emph{cut-off}), the last TLEs of both objects with epoch $\leq$ TCA$\,-\,L$ are taken and propagated without manoeuvres up to the TCA. We define:
\begin{itemize}
\item $\Delta\mathbf r_{\mathrm{obs}}(\mathrm{TCA})$: the \emph{observed} relative position at the TCA, obtained with causal TLE selection, which incorporates all manoeuvres that occurred;
\item $\Delta\mathbf r_{\mathrm{bal}}^{(L)}(\mathrm{TCA})$: the \emph{ballistic} relative position at the TCA, obtained by propagating the TLEs of cut-off $L$ without manoeuvres;
\item $\delta_L$: the discrepancy between the two,
\end{itemize}
\begin{equation}
\delta_L = \left\lVert \Delta\mathbf r_{\mathrm{bal}}^{(L)}(\mathrm{TCA}) - \Delta\mathbf r_{\mathrm{obs}}(\mathrm{TCA})\right\rVert .
\end{equation}
If neither object manoeuvred after the cut-off, $\delta_L$ only reflects the propagation error. The expected uncertainty combines in quadrature two terms from Table~\ref{tab:error}: the relative error for the actual lead of the oldest TLE in the cut-off, $L_{\mathrm{ef}}$, and the error of the observed position itself, whose TLEs are $L_{\mathrm{obs}}$ old at the TCA:
\begin{equation}
\varepsilon_{95} = \sqrt{\varepsilon_{95}^{\mathrm{rel}}(L_{\mathrm{ef}})^2 + \varepsilon_{95}^{\mathrm{rel}}(L_{\mathrm{obs}})^2}.
\end{equation}
H0 is rejected for that cut-off if $\delta_L > K\,\varepsilon_{95}$, with $K = 3$.

The value of $K$ was set so that rejection requires a discrepancy larger than any observed without manoeuvres. The empirical error distribution is not Gaussian and its tail extends beyond the 95th percentile: in the reference period (5087 manoeuvre-free TLE pairs, all lead times and both objects), the maximum error was 2.25~$\varepsilon_{95}$, 4 pairs exceeded $2\,\varepsilon_{95}$ and none exceeded $3\,\varepsilon_{95}$. With $K = 3$, therefore, no propagation of the reference period would have rejected H0. The influence of this choice is assessed in Section~\ref{sec:sensibilidad}.

\paragraph{R3, direction of the effect.} The total time within 5~km in TCA$\,\pm\,$2~days (sum of all intervals, continuous or not) is compared between the observed trajectory and the manoeuvre-free one. Only cut-offs \emph{with power} are evaluated, i.e. those in which the manoeuvre-free minimum distance, after subtracting $K\varepsilon_{95}$, stays above 5~km; in the others, the propagation error prevents a decision.

\paragraph{R2, attribution.} For each cut-off, \emph{only one} of the objects is propagated without manoeuvres and the other is taken as observed. The manoeuvres of an object are considered necessary if, without them, the relative position at the TCA departs from the observed one by more than $K$ times the error of that object. This test is symmetric and does not depend on the thresholds of the manoeuvre detector.

H0 is rejected and the approach is attributed to an object if R1 is met, R3 is met in all cut-offs with power and R2 identifies the object in all of them.

\section{Results}

\subsection{Orbital history of each object}\label{sec:historia}

\begin{figure}[H]
\centering
\includegraphics[width=0.85\textwidth]{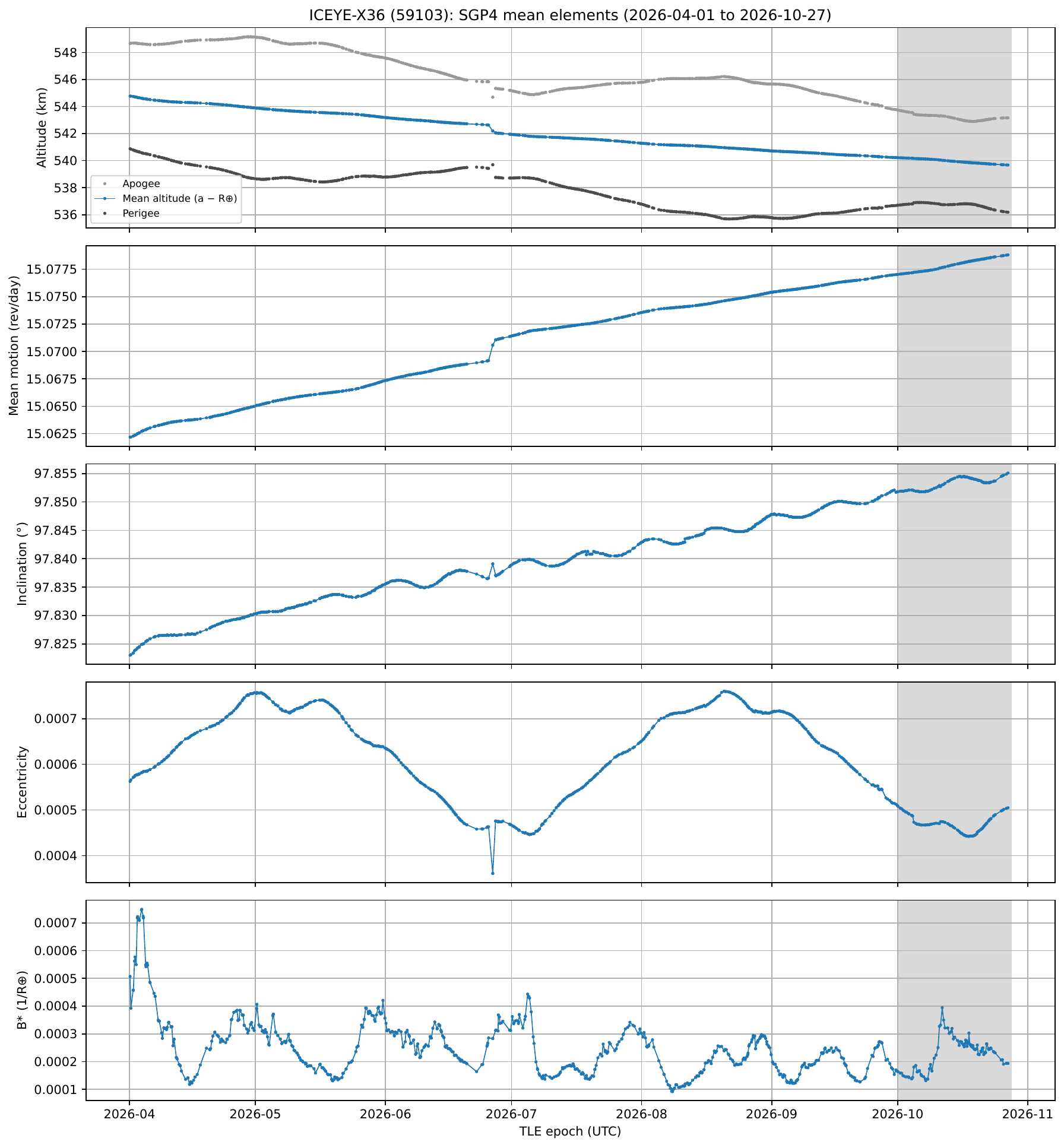}
\caption{SGP4 mean elements of \Ice{} (NORAD 59103), April--October 2026. The grey band marks the analysis window.}
\label{fig:hist59103}
\end{figure}

\Ice{} maintained a near-circular orbit at $\approx$540~km mean altitude, with an inclination of 97.84$^\circ$ and a slow decay (Figure~\ref{fig:hist59103}). The detector flagged 23 events over the whole history, all of very small magnitude (gross $\Delta v$: median 0.03~m/s, maximum 0.68~m/s); their frequency in September--October (6 events) matches that of the reference period (6.7 expected, $p = 0.66$). With a threshold $z > 20$ there is no event of \Ice{} in September--October. Its behaviour is consistent with TLE noise or minor adjustments, and shows no changes during the encounter.

\begin{figure}[H]
\centering
\includegraphics[width=0.85\textwidth]{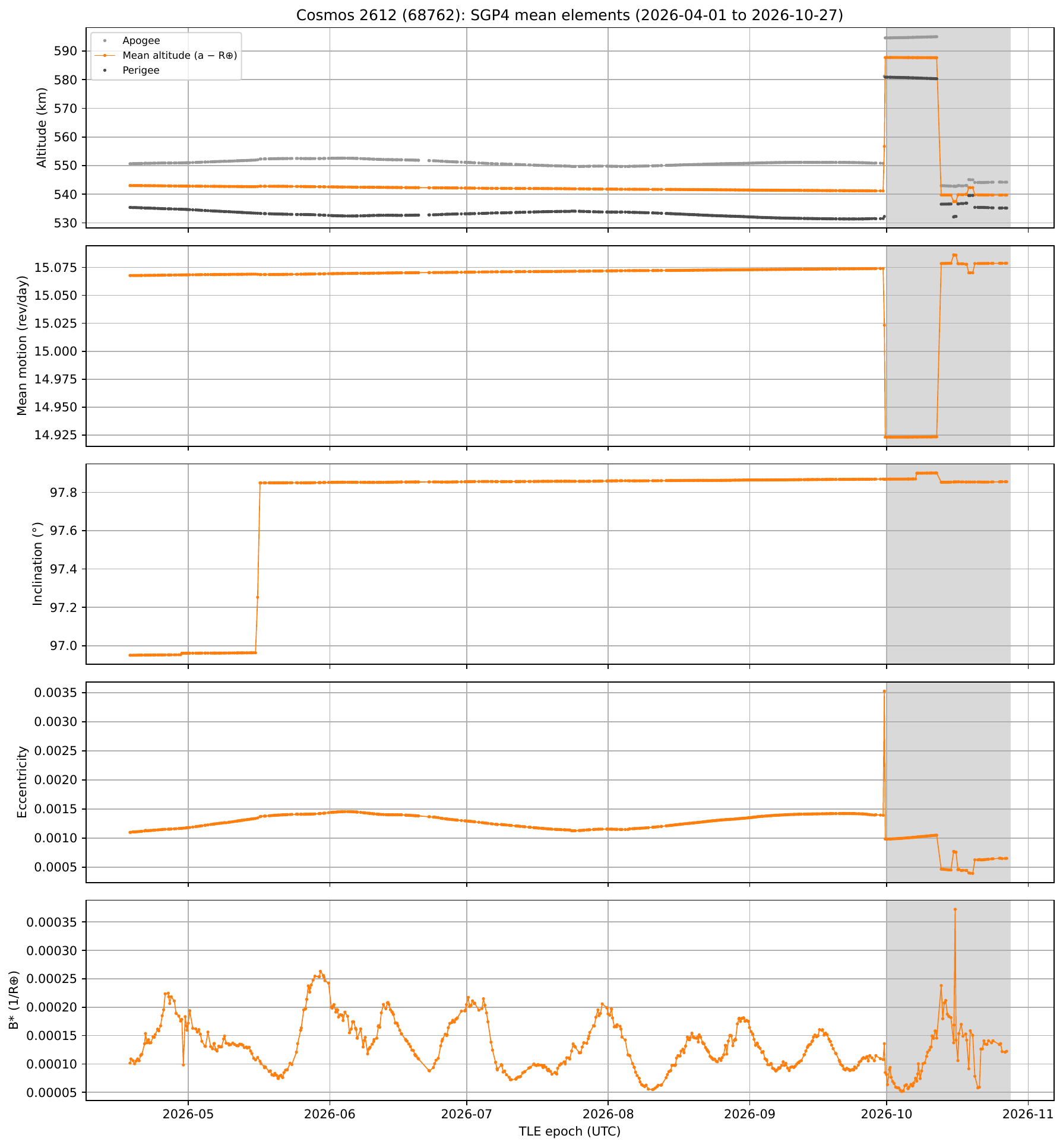}
\caption{SGP4 mean elements of \Cos{} (NORAD 68762), April--October 2026.}
\label{fig:hist68762}
\end{figure}

\Cos{}, in contrast, shows large orbital changes (Figure~\ref{fig:hist68762}, Table~\ref{tab:maniobras}). On 15--16 May its inclination changed from 96.96$^\circ$ to 97.86$^\circ$, practically that of \Ice{} (97.84$^\circ$). On 30 September its mean altitude rose from 541 to 588~km; an intermediate TLE with eccentricity 0.0035 and mean altitude of 557~km is consistent with a two-impulse transfer. Between 10 and 12 October it descended again to the altitude of \Ice{}, with a small inclination adjustment, and between 15 and 21 October the detector flags 11 of the 18 pairs of consecutive TLEs, with scores up to $z \approx 1400$.

The last panel of Figures~\ref{fig:hist59103} and~\ref{fig:hist68762} shows the drag term $B^*$ of each TLE. In both objects it varies from one TLE to the next without a defined trend, and for \Cos{} it shows isolated extreme values during the approach and the retreat. This parameter is not used as an indicator in the analysis for two reasons. First, its variation between consecutive TLEs is large, and this work has no calibrated criterion to distinguish a manoeuvre from fitting noise in it; defining one after knowing the results would introduce a bias. Second, its effect is already included in the test: SGP4 propagates each TLE with its $B^*$, so the manoeuvre-free propagations incorporate the differential drag between both objects, and the null hypothesis covers that cause of natural drift.

\begin{table}[H]
\centering
\caption{Manoeuvre events of \Cos{} with $z > 20$ since May 2026. $\Delta v$ in m/s (lower bounds). If all events detected since May ($|z| > 5$) are included, the totals are 174.0 (net) and 179.7~m/s (gross); hence the range of 174--180~m/s quoted in the text.}
\label{tab:maniobras}
\resizebox{\textwidth}{!}{%
\begin{tabular}{llrrrrr}
\toprule
Start -- end (UTC, 2026) & Interpretation & $\Delta a$ (km) & $\Delta i$ ($^\circ$) & Net $\Delta v$ & Gross $\Delta v$ & $z_{\max}$ \\
\midrule
15-05 17:11 -- 16-05 17:05 & Plane match & $+0.14$ & $+0.886$ & 117.4 & 117.4 & 4025 \\
30-09 05:09 -- 30-09 16:22 & Raise to drift orbit & $+46.54$ & $-0.001$ & 25.5 & 25.5 & 19\,111 \\
07-10 07:41 -- 07-10 14:07 & Inclination adjustment & $\approx 0$ & $+0.030$ & 4.0 & 4.0 & 204 \\
10-10 17:45 -- 12-10 22:39 & Lowering and orbit matching & $-47.87$ & $-0.048$ & 26.8 & 26.8 & 22\,021 \\
15-10 01:37 -- 21-10 15:17 & Approach and retreat & $+0.03$ & $+0.000$ & 0.03 & 5.7 & 1420 \\
\midrule
\multicolumn{3}{l}{Total} & & 173.7 & 179.4 & \\
\bottomrule
\end{tabular}}
\end{table}

\subsection{Relative geometry and timeline of the encounter}

\begin{figure}[H]
\centering
\includegraphics[width=0.85\textwidth]{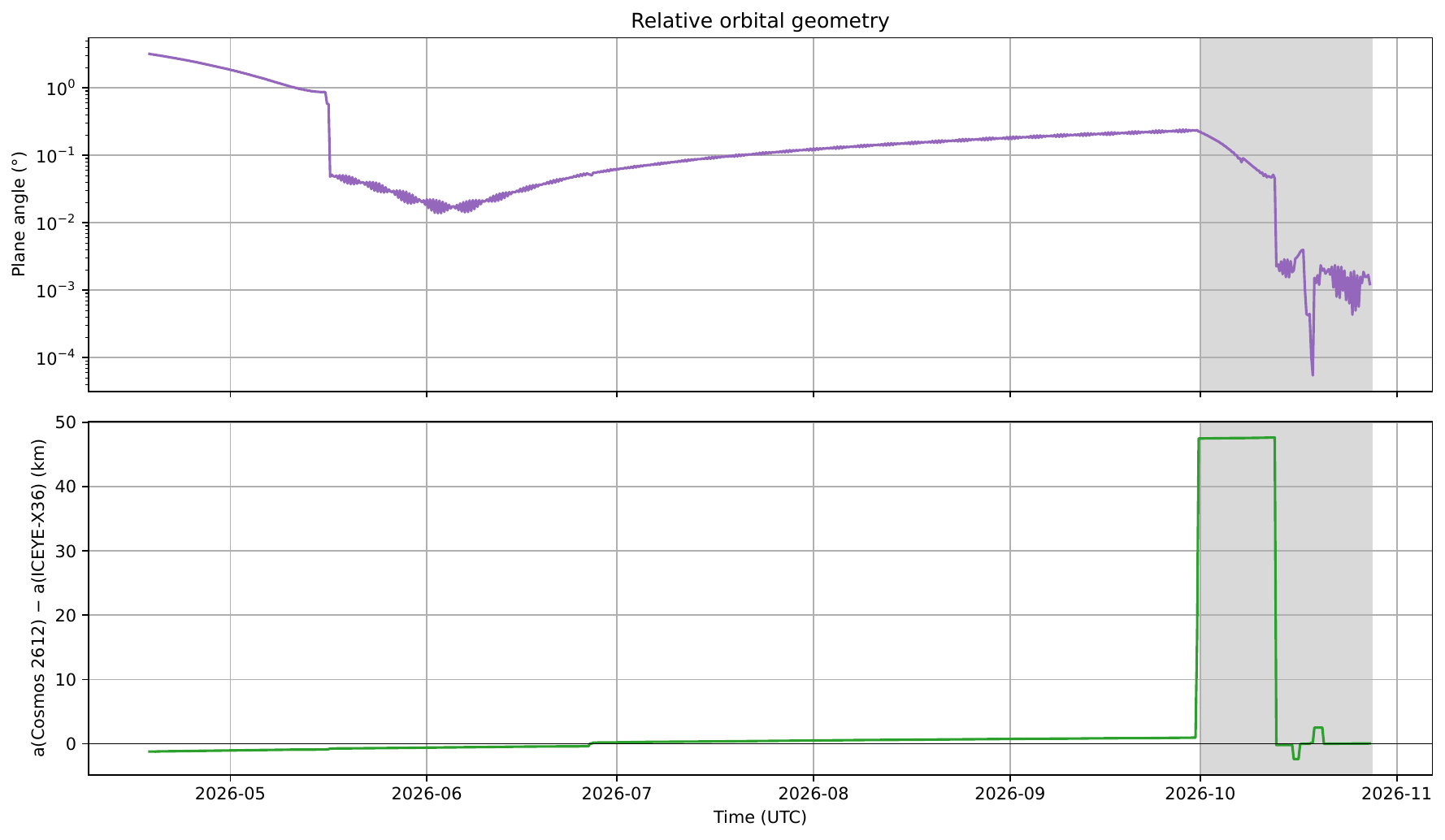}
\caption{Angle between the orbital planes (top, logarithmic scale) and difference in semi-major axes $a_{68762} - a_{59103}$ (bottom).}
\label{fig:geometria}
\end{figure}

Figure~\ref{fig:geometria} shows the convergence of the orbits. In mid-April the angle between the orbital planes was $\approx$3$^\circ$ and it decreased without manoeuvres to $\approx$0.9$^\circ$ by mid-May, consistent with the differential nodal precession associated with the different inclinations: the secular drift of the ascending node due to $J_2$, $\dot\Omega = -\tfrac32 J_2\, n\,(R_\oplus/p)^2\cos i$, where $J_2$ is the coefficient associated with the Earth's oblateness and $p = a(1-e^2)$ the semi-latus rectum, depends on the inclination and the semi-major axis \cite{gurfil2004}. The May manoeuvre removed the inclination difference and left the angle at $\approx$0.05$^\circ$. Between June and September the angle grew slowly to $\approx$0.23$^\circ$; during the drift phase at higher altitude it decreased again to $\approx$0.06$^\circ$, and after the October adjustments it was $\approx$0.003$^\circ$. At the TCA the angle between the planes was 0.0004$^\circ$ and the difference in semi-major axes was $-2$~m.

\begin{figure}[H]
\centering
\includegraphics[width=0.85\textwidth]{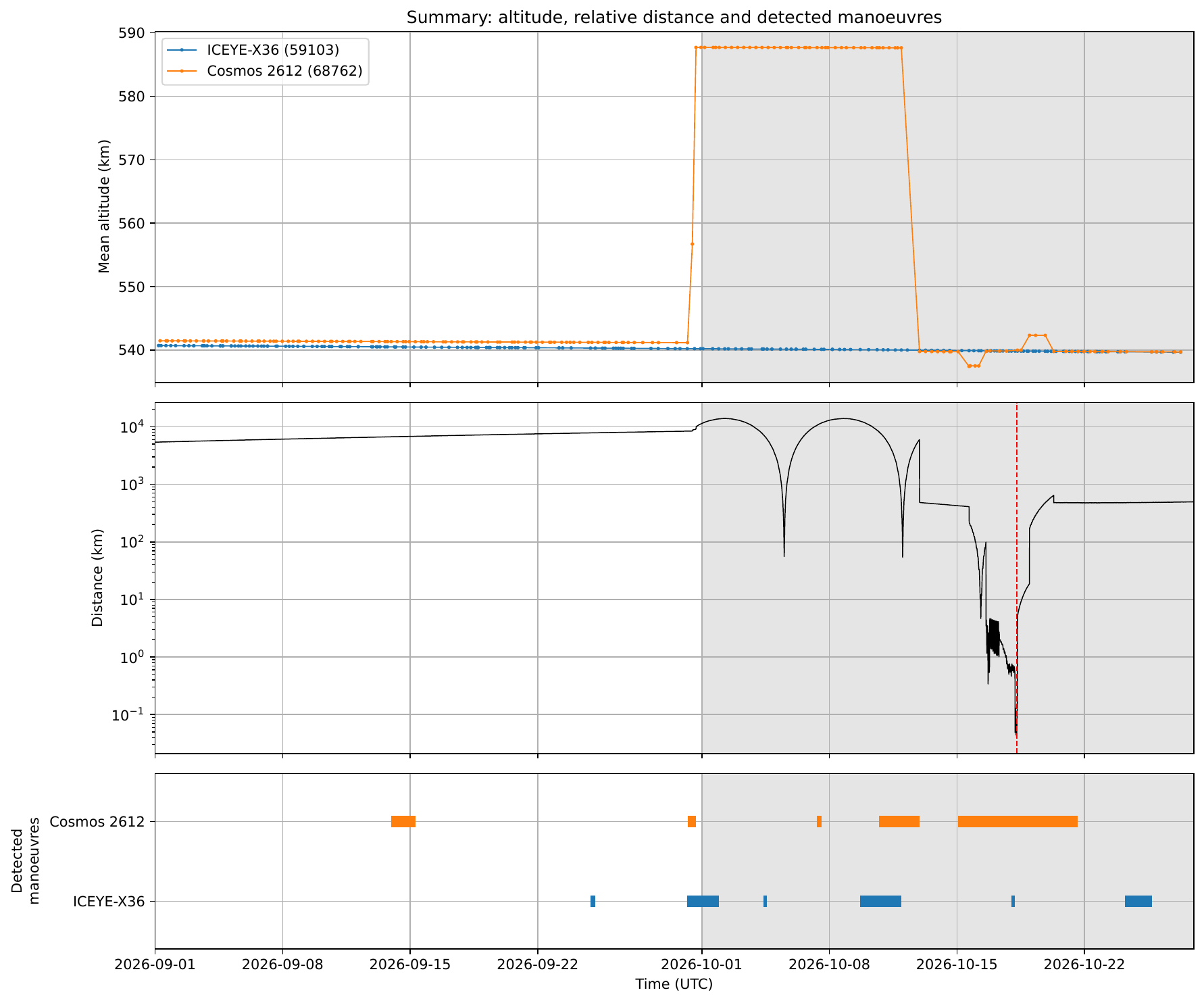}
\caption{Summary from 1 September to 27 October 2026: mean altitude of both objects (top), relative distance on a logarithmic scale with the TCA in red (middle) and events detected by the manoeuvre algorithm for each object (bottom). The events of \Ice{} correspond to changes of very small magnitude (Section~\ref{sec:historia}).}
\label{fig:sintesis}
\end{figure}

Figure~\ref{fig:sintesis} summarises the sequence, which is described in five phases:

\begin{enumerate}
\item \textbf{Phasing drift (30 September -- 12 October).} With \Cos{} 47.5~km higher, the difference in mean motion (medians of 15.077 and 14.923 rev/day) implies a synodic period of $1/\Delta n \approx 6.50$~days. Two passes are observed, on 5 October at 12:28~UTC (54.9~km) and on 12 October at 00:30~UTC (53.9~km), 6.50~days apart. The local relative velocity during this phase had a median of $\approx$79~m/s, in agreement with the expected along-track drift, $\tfrac32 n\,\Delta a \approx 78$~m/s \cite{seong2013}.
\item \textbf{Far parking (13--14 October).} After the lowering, \Cos{} was on the orbit of \Ice{}, $\approx$454~km behind in the along-track direction, with a local relative velocity of $\approx$0.7~m/s.
\item \textbf{Approach (15--16 October).} The separation decreased from $\approx$430~km to less than 5~km. The mean altitude of \Cos{} dropped $\approx$2.4~km relative to \Ice{}, which allowed it to catch up.
\item \textbf{Proximity (16--18 October).} The separation stayed below 5~km for 41.4 continuous hours (16 October 14:15 -- 18 October 07:38~UTC) and below 1~km for 14.6~h (17 October 17:04 -- 18 October 07:38~UTC).
\item \textbf{Retreat (18--20 October).} \Cos{} raised its mean altitude $\approx$2.5~km above that of \Ice{} and drew back to $\approx$480~km on 20 October. Until 23 October the separation remained at $\approx$478~km and then increased by $\approx$4~km/day, reaching $\approx$494~km at the end of 27 October.
\end{enumerate}

\subsection{Closest approach}

\begin{figure}[H]
\centering
\includegraphics[width=0.85\textwidth]{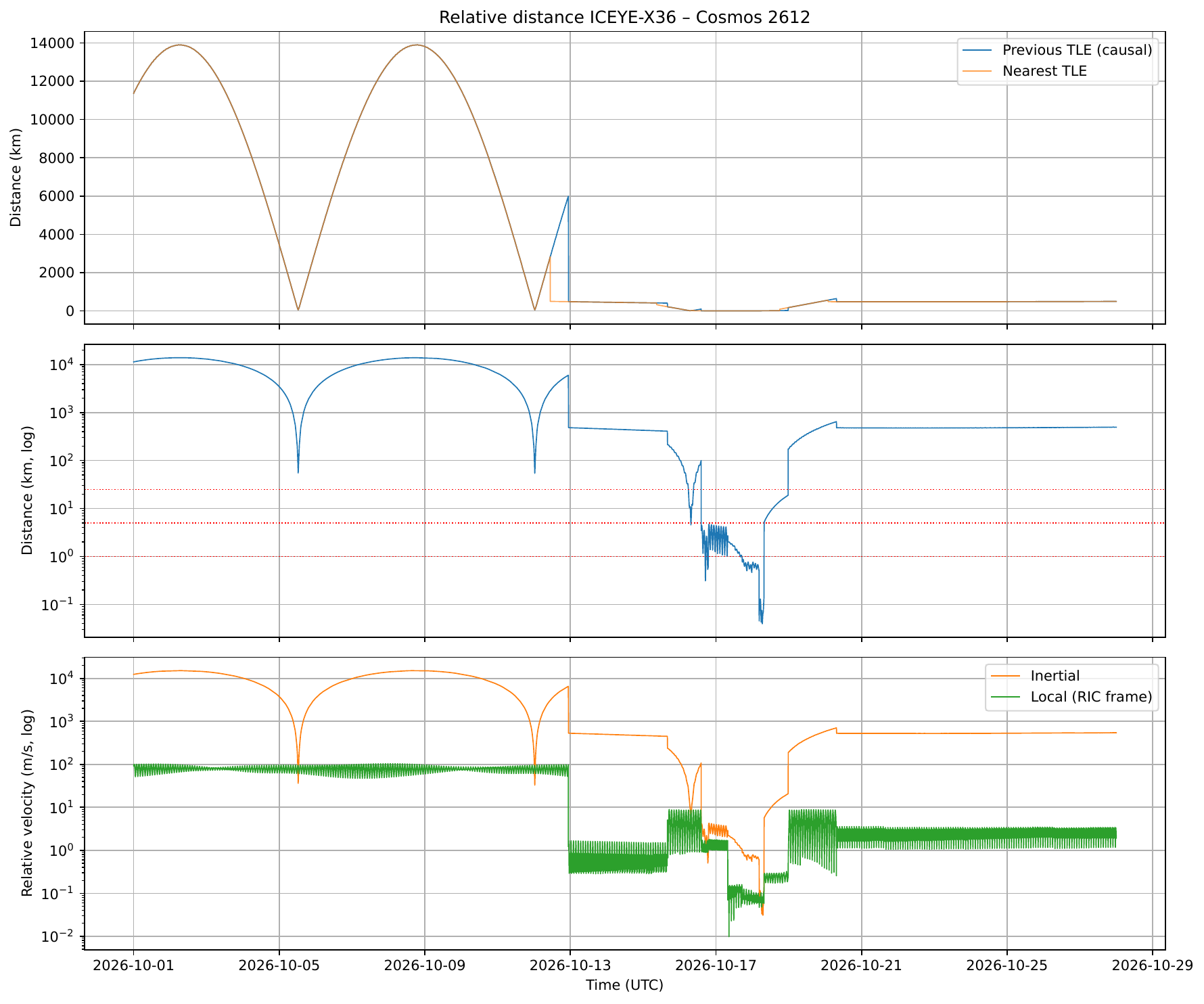}
\caption{Relative distance from 1 to 27 October on a linear scale (top) and logarithmic scale with thresholds of 1, 5 and 25~km (middle), and inertial and local relative velocity (bottom). The discontinuities correspond to TLE changes after a manoeuvre.}
\label{fig:distancia}
\end{figure}

\begin{figure}[H]
\centering
\includegraphics[width=0.85\textwidth]{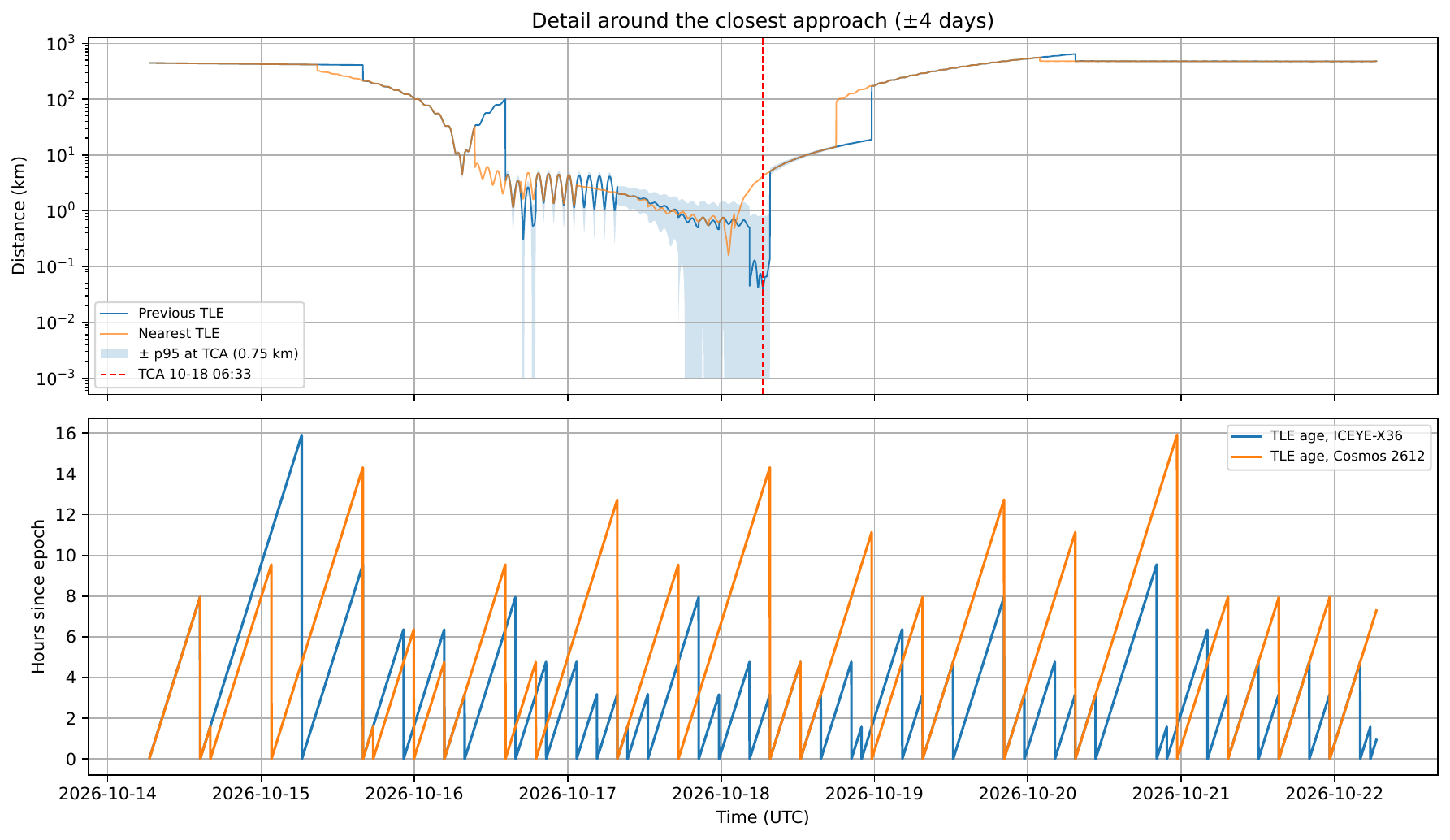}
\caption{Detail of $\pm$4 days around the TCA with the p95 uncertainty band (top) and age of the TLEs used (bottom).}
\label{fig:detalle}
\end{figure}

The TCA, refined to a resolution of 1~s, occurred on \textbf{18 October 2026 at 06:33:14~UTC} (Table~\ref{tab:tca}, Figures~\ref{fig:distancia} and~\ref{fig:detalle}). The computed distance (39~m) is smaller than the uncertainty of the method (0.75~km, 95th percentile or p95, with TLEs 2.1 and 13.2~h old), so it can only be stated that the separation was \textbf{below 0.79~km} with 95\,\% confidence. Published estimates of a few tens of metres, such as the $\approx$26~m quoted by The Washington Times \cite{washingtontimes2026}, are compatible with this result, but cannot be confirmed or refuted with GP data.

\begin{table}[H]
\centering
\caption{Characteristics of the approach.}
\label{tab:tca}
\begin{tabular}{lr}
\toprule
TCA (UTC) & 18-10-2026 06:33:14 \\
Minimum separation (p95 bound) & $< 0.79$~km \\
Local relative velocity at the TCA & 0.08~m/s \\
Time within $< 1$ / $< 5$ / $< 25$~km (longest run) & 14.6 / 41.4 / 57.3~h \\
Median local velocity within $< 1$~km (p90) & 0.08 (0.13)~m/s \\
Median local velocity within $< 5$~km & 0.13~m/s \\
Angle between planes at the TCA & 0.0004$^\circ$ \\
$a_{68762} - a_{59103}$ at the TCA & $-2$~m \\
\bottomrule
\end{tabular}
\end{table}

\subsection{Relative kinematics}

\begin{figure}[H]
\centering
\includegraphics[width=\textwidth]{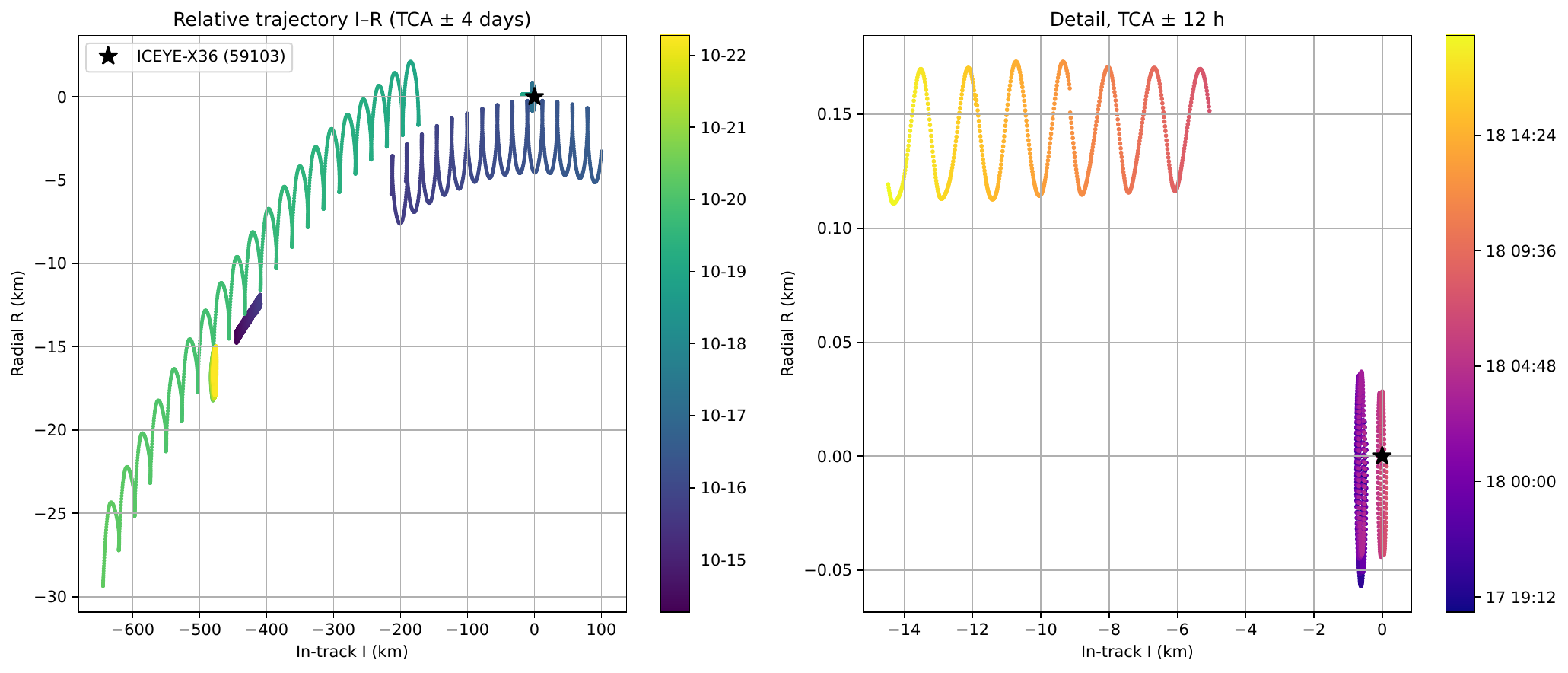}
\caption{Trajectory of \Cos{} in the along-track--radial plane of the local frame of \Ice{} (star) for $\pm$4 days (left) and $\pm$12~h (right) around the TCA. Positions below $\approx$0.8~km cannot be resolved.}
\label{fig:ric}
\end{figure}

In the local frame of \Ice{} (Figure~\ref{fig:ric}), \Cos{} remained behind in the along-track direction from the far parking to the end of the period, at distances of up to $\approx$500~km, except for positions below the resolution of the data. The exceptions are the intervals in which the causal series propagates a TLE prior to a manoeuvre and the next TLE contradicts it. The most notable occurs between the epochs of the \Cos{} TLEs of 16 October at 04:40 and 14:14~UTC: the first predicts that \Cos{} overtakes \Ice{} and reaches $\approx$100~km ahead, but the second places it $\approx$3~km behind. Similarly, on 20 October the causal series places \Cos{} up to $\approx$640~km behind before the TLE of 07:26~UTC, which brings it back to $\approx$480~km. These segments are not interpreted as real motion. The approach took place along the orbital track, from behind. The cross-track component stayed below 0.5~km, so the encounter was essentially coplanar. The mean radial offset of $\approx$$-15$~km observed at 450~km separation corresponds to the curvature of the orbit ($\approx I^2/2r$) and not to an altitude difference. Relative positions below one kilometre, including possible station-keeping positions, are below the resolution of the data and are not interpreted.

\subsection{Hypothesis test and attribution}

\begin{figure}[H]
\centering
\includegraphics[width=0.85\textwidth]{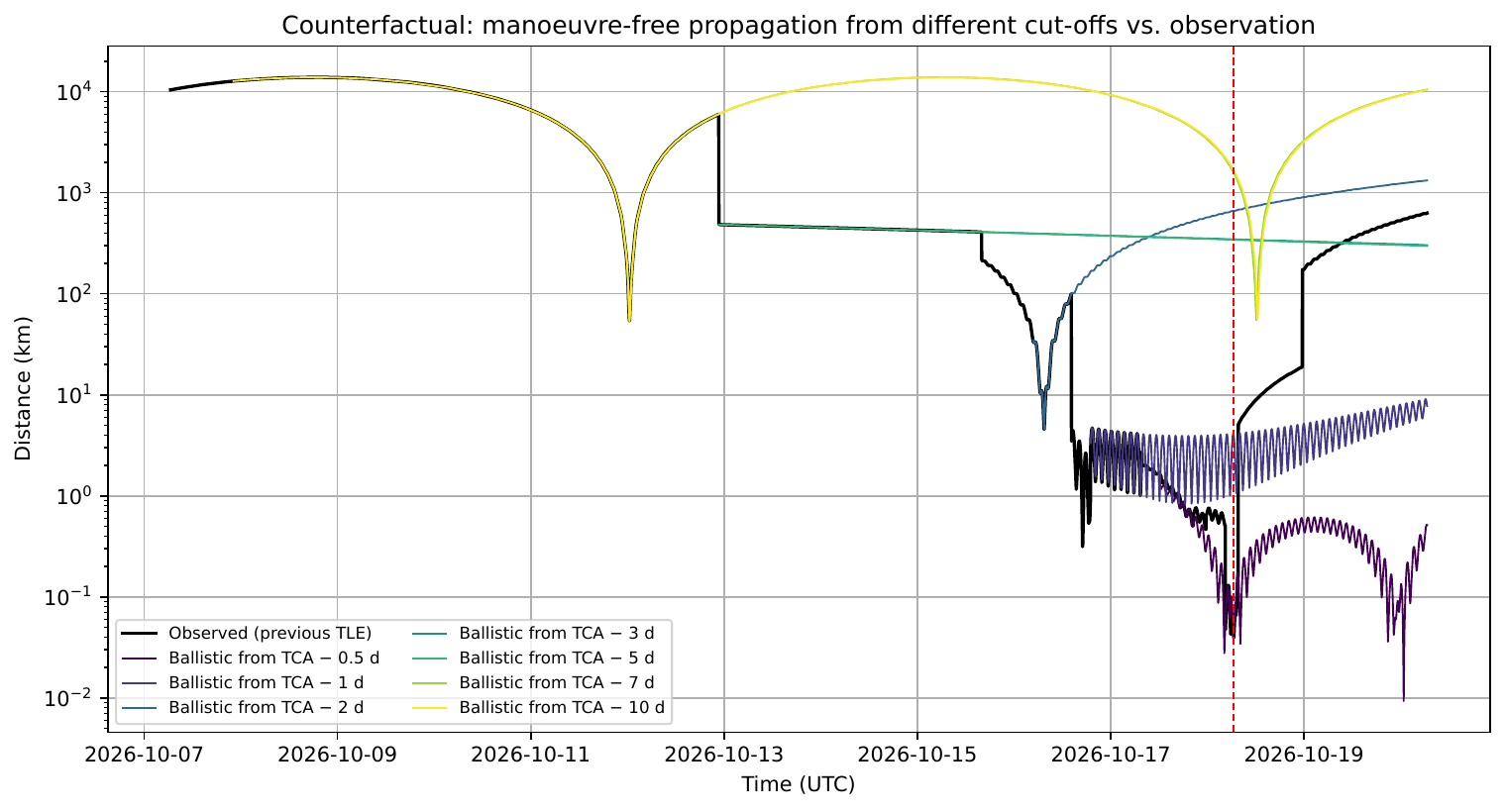}
\caption{Observed distance (black) and manoeuvre-free propagations from different cut-offs before the TCA (dashed red line).}
\label{fig:contrafactual}
\end{figure}

\begin{figure}[H]
\centering
\includegraphics[width=0.8\textwidth]{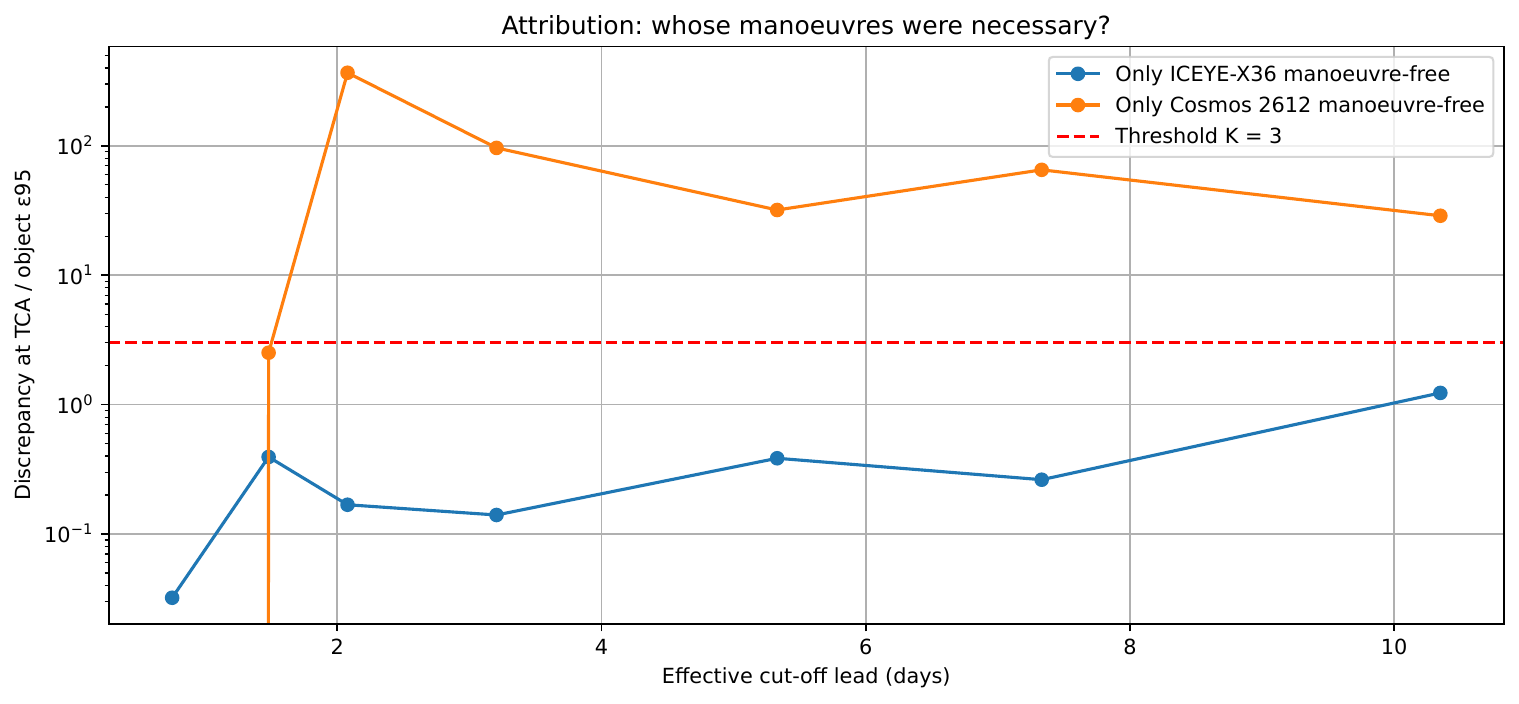}
\caption{Attribution: discrepancy at the TCA, normalised by the p95 error of each object, when only the manoeuvres of \Ice{} (blue) or only those of \Cos{} (orange) are removed. The red line marks the threshold $K = 3$.}
\label{fig:atribucion}
\end{figure}

Table~\ref{tab:contraste} and Figures~\ref{fig:contrafactual} and~\ref{fig:atribucion} summarise the test. For the 0.75 and 1.48~day cut-offs, the manoeuvre-free propagation reproduces the observation at the TCA; this serves as a check of the method. From the 2.08~day cut-off onwards, H0 is rejected in every case, with discrepancies of 16 to 195 times the expected error. The 3.21 and 5.33~day cut-offs have power to assess the direction of the effect: without manoeuvres, the separation at the TCA would have been $\approx$340~km and the time within 5~km zero, compared with the 41.6~h observed in total (41.4~h continuous). In the 7.33 and 10.35~day cut-offs, the manoeuvre-free propagation predicts the next natural pass of the phasing drift on 18 October at around 12:10--12:27~UTC, at $\approx$55~km, but its error (44--99~km) prevents their use to assess the direction.

The attribution is clear in all cut-offs: if only the manoeuvres of \Cos{} are removed, the discrepancy is 29 to 367 times its error in every cut-off that rejects H0; if only those of \Ice{} are removed, the discrepancy does not exceed 1.2 times its error in any cut-off. Moreover, in the short cut-offs the manoeuvre-free propagation would have kept the objects close for 75--95~h in total, compared with the 41.6~h observed, which indicates that the retreat was also manoeuvred.

\begin{table}[H]
\centering
\caption{Counterfactual test. $d_{\mathrm{bal}}$: separation at the TCA without manoeuvres; $\delta/\varepsilon_{95}$: normalised discrepancy; $h_{<5}$: hours within 5~km without manoeuvres in TCA$\,\pm\,$2~d (observed: 41.6~h in total, 41.4~h continuous); attribution columns: normalised discrepancy when only the manoeuvres of each object are removed.}
\label{tab:contraste}
\resizebox{\textwidth}{!}{%
\begin{tabular}{rrrrcccrr}
\toprule
Cut-off (d) & $d_{\mathrm{bal}}$ (km) & $\varepsilon_{95}$ (km) & $\delta/\varepsilon_{95}$ & R1 & Power & $h_{<5}$ (h) & Only \Ice{} & Only \Cos{} \\
\midrule
0.75  & 0.04   & 1.24  & 0.03  & --   & --   & 95.3 & 0.03 & 0.00 \\
1.48  & 3.97   & 2.26  & 1.75  & --   & --   & 75.2 & 0.39 & 2.52 \\
2.08  & 659    & 3.39  & 194.6 & Yes & --   & 0.15 & 0.17 & 367.1 \\
3.21  & 345    & 7.48  & 46.1  & Yes & Yes & 0    & 0.14 & 96.4 \\
5.33  & 342    & 21.4  & 15.9  & Yes & Yes & 0    & 0.38 & 31.9 \\
7.33  & 1564   & 43.9  & 35.7  & Yes & --   & 0    & 0.26 & 65.2 \\
10.35 & 1646   & 98.5  & 16.7  & Yes & --   & 0    & 1.23 & 28.8 \\
\bottomrule
\end{tabular}}
\end{table}

\section{Robustness and bias control}

\subsection{Parameter sensitivity}\label{sec:sensibilidad}

The full evaluation was repeated for $K \in \{2, 3, 4\}$, proximity threshold $\in \{1, 5, 25\}$~km and error percentile $\in \{95, 99\}$, 18 combinations in total. All 18 give the same result: H0 rejected and approach attributed to \Cos{}; in none of them are necessary manoeuvres attributed to \Ice{}. Since the smallest normalised discrepancy among the cut-offs rejecting H0 is 15.9 (Table~\ref{tab:contraste}), R1 would still be met in all of them for any $K < 15.9$. The conclusion rests on two cut-offs with power (3.21 and 5.33~days), although the attribution also holds in all other cut-offs that reject H0.

\subsection{Negative controls}

Exactly the same procedure was applied to all other local distance minima below 100~km in the analysis window (Table~\ref{tab:controles}). The pass of 12 October does not reject H0: it is reproduced by the manoeuvre-free propagation. The pass of 5 October rejects H0 in the cut-offs before 30 September, because it is a consequence of the raising manoeuvre, but it is not classified as a proximity event, since there was no time spent within 5~km. The method thus distinguishes a fast pass, even if caused by an earlier manoeuvre, from a sustained approach. In addition, the same procedure was applied to synthetic scenarios with known truth, generated with SGP4 and with noise in the elements (Table~\ref{tab:sinteticos}). In four natural passes without manoeuvres, with minimum separations of 0.15--0.26~km and about 10~h within 5~km, the verdict was \emph{H0 not rejected}. In three approaches induced by manoeuvres of a single object (plane match, phasing drift, far parking, approach, hold and retreat), the approach was attributed to that object and not to the other. Because the noise of the synthetic elements is smaller than that of real TLEs and the empirical error tables of Table~\ref{tab:error} are used, this check verifies the logic of the decision rules, not the error model.

\begin{table}[H]
\centering
\caption{Synthetic-data check. Object A (target) does not manoeuvre; object B manoeuvres only in the induced scenarios.}
\label{tab:sinteticos}
\begin{tabular}{llrrl}
\toprule
Scenario & TCA (UTC, 2026) & $d_{\min}$ (km) & $h_{<5}$ (h) & Verdict \\
\midrule
Natural  & 06-10 05:58 & 0.25 & 10.6 & H0 not rejected \\
Natural  & 12-10 06:05 & 0.26 & 10.7 & H0 not rejected \\
Natural  & 18-10 06:12 & 0.25 & 10.2 & H0 not rejected \\
Natural  & 24-10 05:32 & 0.15 & 10.0 & H0 not rejected \\
Induced & 12-10 12:28 & 0.06 & 33.2 & Induced by B \\
Induced & 18-10 01:26 & 0.02 & 33.2 & Induced by B \\
Induced & 22-10 23:16 & 0.02 & 32.4 & Induced by B \\
\bottomrule
\end{tabular}
\end{table}

\begin{table}[H]
\centering
\caption{Negative controls: local approaches of the same pair evaluated with the same procedure. $h_{<5}$: total hours within 5~km in $\pm$2~days around each approach; $v_{\mathrm{loc}}$ at the instant of the minimum.}
\label{tab:controles}
\resizebox{\textwidth}{!}{%
\begin{tabular}{lrrrl}
\toprule
Date (UTC) & $d_{\min}$ (km) & $v_{\mathrm{loc}}$ (m/s) & $h_{<5}$ (h) & Result \\
\midrule
05-10 12:28 & 54.9 & 93.9 & 0 & Does not increase proximity \\
12-10 00:30 & 53.9 & 91.3 & 0 & H0 not rejected \\
18-10 06:33 & $< 0.79$ & 0.08 & 41.6 & Induced by \Cos{} \\
\bottomrule
\end{tabular}}
\end{table}

\subsection{Statement of changes to the decision rules}\label{sec:cambios}

The decision rules were fixed before running the analysis on real data and, after the first run, two of them were revised. This section describes the initial version, what was observed when applying it and the modification adopted.

Initially, R3 required the observed minimum distance in TCA$\,\pm\,$2~days to be smaller than the ballistic one, after subtracting $K\varepsilon_{95}$, in \emph{all} cut-offs rejecting H0, and R2 attributed the approach according to the presence of manoeuvre-detector events with $|z| > 5$ between the cut-off and the TCA. With these rules R1 was met (H0 rejected from the 2.08~day cut-off onwards), but the overall verdict was \emph{not attributable}.

Examining the tests revealed the following:
\begin{itemize}
\item R3 was not met in the 2.08, 7.33 and 10.35~day cut-offs. In all of them the ballistic minimum distance (4.5, 54.8 and 56.3~km) was smaller than $K\varepsilon_{95}$ (10, 132 and 296~km), so the comparison fell below the resolution of the data and could not have been favourable in any scenario.
\item Between the cut-off and the TCA the detector flagged events in both objects. Those of \Ice{} occur at the same rate as in the reference period ($p = 0.66$) and are of very small magnitude (Section~\ref{sec:historia}), consistent with TLE noise, so the presence of an event was not enough to attribute the approach.
\end{itemize}

R3 was therefore restricted to the cut-offs with power, defined by a criterion that depends only on the propagation error and the ballistic trajectory, and R2 was replaced by the per-object counterfactual test described in Section~\ref{sec:contraste}, which does not depend on the detector. R1 was not modified: the conclusion that the approach required manoeuvres is the same with the initial and the revised rules. The revision affects the direction of the effect and the attribution to a specific object.

Because the revision was made after the first result was known, the attribution does not have the value of a fully pre-specified test. It is therefore complemented by the sensitivity analysis (Section~\ref{sec:sensibilidad}), the negative controls and the synthetic-data check, and both verdicts are kept in the reproducible material.

\section{Discussion}

\subsection{Nature of the encounter}

The results rule out that the approach was due to the natural evolution of the orbits. The observed sequence (plane match, phasing drift from a higher orbit, orbit matching at a distance, approach from behind along the track, proximity and controlled retreat) is consistent with the definition of RPO: a series of manoeuvres that bring two objects together at a planned time and place and keep one of them on a relative trajectory for a given time \cite{samson2026}. The manoeuvres that made the encounter possible were performed by \Cos{}; those of \Ice{} are not needed to explain it, and within the resolution of the data this object behaved passively throughout the period. The minimum cumulative $\Delta v$ since May, 174--180~m/s, is dominated by the plane match (117~m/s, 65--67\,\% of the total).

The evolution of the angle between the planes suggests a deliberate use of differential nodal precession: the angle decreased without manoeuvres between April and May, and again during the drift at higher altitude, which would have reduced the $\Delta v$ cost of aligning the planes. This interpretation is consistent with the data, but it is not demonstrated in this work.

\subsection{Comparison with public information}

The published minimum distances are compatible with the results. The LeoLabs estimate of $\approx$500~m \cite{cnn2026} and the Scientific American estimate of a few hundred metres \cite{randolph2026} fall within the 0.79~km bound obtained here. AeroTime places the approach at around 01:35~UTC \cite{aerotime2026}; if this refers to 18 October, the distance computed in this analysis at that instant is 0.65~km, within the continuous interval below 1~km (17 October 17:04 -- 18 October 07:38~UTC). Estimates of a few tens of metres \cite{washingtontimes2026} are below the resolution of GP data. AeroTime's conclusion that the uncertainty of public data prevents a reliable calculation of the minimum separation \cite{aerotime2026} agrees with the limitation quantified in this work.

Other claims cannot be assessed with kinematic data. The \emph{non-cooperative} character (without the operator's permission) refers to agreements and communications, not to trajectories: a consented approach would produce the same orbital signature, and the absence of manoeuvres by \Ice{} is compatible with lack of awareness or of margin to manoeuvre as well as with acceptance. Intent, whether inspection, interference or other, cannot be determined either. There are proposals to infer intent from the behaviour observed during an RPO \cite{kelecy2022}, but applying them is beyond the scope of this work, and the characterisation of the event as deliberate and unsolicited \cite{randolph2026} is an interpretation that kinematic data can neither confirm nor rule out. Likewise, there is no orbital evidence of contact with or disturbance of \Ice{}, but only its operator can confirm its status.

\subsection{Limitations}

\begin{itemize}
\item \textbf{Data accuracy.} The error of propagated TLEs ranges from hundreds of metres to kilometres (Table~\ref{tab:error}); the minimum distance and the geometry cannot be resolved below $\approx$0.8~km.
\item \textbf{No covariance.} TLEs include no formal uncertainty, so no probability of collision is computed; the uncertainty used is empirical.
\item \textbf{Manoeuvre resolution.} Manoeuvres can only be located between consecutive TLE epochs, and the $\Delta v$ values are lower bounds.
\item \textbf{Differential drag.} A difference in $B^*$ produces natural drift; SGP4 includes it and its normal variability is part of the empirical error $\varepsilon_{95}$. An abrupt change in attitude or area could resemble a small manoeuvre in the detector, or produce a discrepancy in the manoeuvre-free propagations not captured by $\varepsilon_{95}$; however, the discrepancies observed for the main event (16 to 195 times $\varepsilon_{95}$) are far larger than the variability measured in the reference period.
\item \textbf{Local velocity at large separations.} At hundreds of kilometres the local velocity includes eccentricity and curvature effects and does not represent the actual drift.
\item \textbf{Controls.} There are only two real negative controls and both belong to the same pair of objects; they verify the behaviour of the method but do not allow a false-positive rate to be estimated.
\item \textbf{Rule revision.} The decision rules were revised after the first run (Section~\ref{sec:cambios}), which is stated explicitly.
\end{itemize}

\section{Conclusions}

\begin{enumerate}
\item The October 2026 approach between \Cos{} and \Ice{} is not compatible with natural orbital evolution: the null hypothesis is rejected in all 18 parameter combinations tested.
\item The approach required manoeuvres by \Cos{}; those of \Ice{} are not needed to reproduce the observations.
\item The encounter was the final stage of a sequence that began five months earlier: plane match in May, phasing drift from 30 September, orbit matching on 10--12 October, approach on 15--16 October, proximity on 16--18 October and manoeuvred retreat on 18--20 October, with a minimum $\Delta v$ of 174--180~m/s.
\item The TCA occurred on 18 October 2026 at 06:33~UTC, with a separation below 0.79~km (p95), a local relative velocity of $\approx$0.08~m/s and 41.4~h within 5~km and 14.6~h within 1~km, in a coplanar approach from behind along the orbital track.
\item Public data do not allow the exact minimum distance, the occurrence of contact, the cooperative character of the operation or its purpose to be determined.
\end{enumerate}

As future work, the procedure should be validated with proximity operations whose timeline is public and applied to a large sample of unrelated pairs to estimate its false-positive rate; systematic searches for candidates in the public catalogue \cite{du2026} can provide that sample.

\section*{Data availability}

\begin{sloppypar}
The code and derived results are available at \url{https://github.com/bnbuitragom/cosmos2612-iceye-x36-rpo-analysis}. The complete analysis, including the original and revised decision rules, is in the notebook \texttt{cosmos2612\_\allowbreak iceye-x36\_\allowbreak proximity\_\allowbreak analysis.ipynb}, which uses NumPy, pandas, SciPy, Matplotlib and \texttt{sgp4} \cite{rhodes_sgp4}. The GP data come from Space-Track \cite{spacetrack}. Due to the redistribution restrictions of the Space-Track user agreement, the original element sets are not included in the repository; the script \texttt{src/\allowbreak download\_\allowbreak data.py} downloads them with a free account on the portal and reproduces the dataset used in this work.
\end{sloppypar}

\section*{CRediT authorship contribution statement}

\textbf{Brayan Nicolas Buitrago Montenegro:} Conceptualization, Methodology, Software, Formal analysis, Investigation, Data curation, Validation, Visualization, Writing -- original draft, Writing -- review \& editing.

\section*{Declaration of competing interest}

The author declares no known competing financial interests or personal relationships that could have appeared to influence the work reported in this paper.

\section*{Funding}

This research did not receive any specific grant from funding agencies in the public, commercial, or not-for-profit sectors.

\appendix
\section{Supplementary figures}
\setcounter{figure}{0}
\renewcommand{\thefigure}{A.\arabic{figure}}

\begin{figure}[H]
\centering
\includegraphics[width=0.8\textwidth]{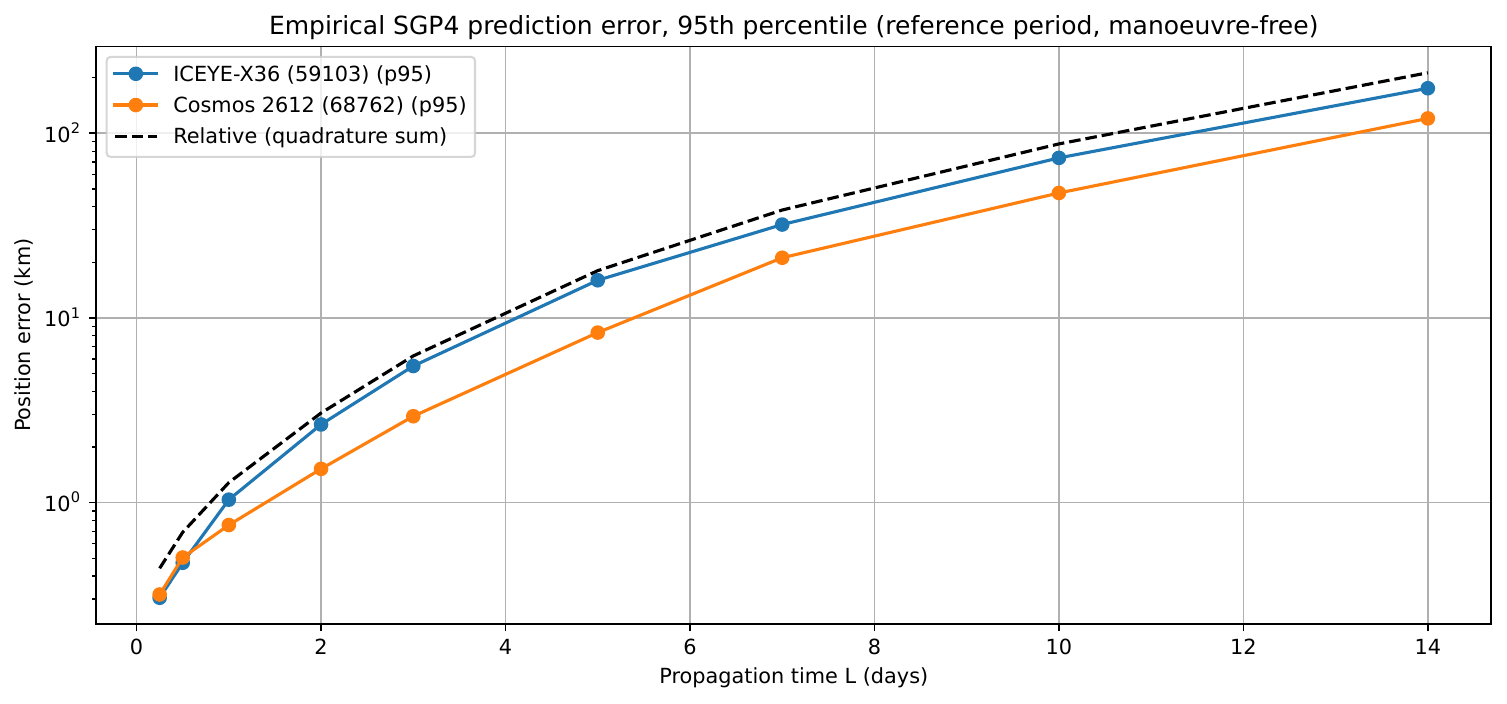}
\caption{Empirical SGP4 prediction error (95th percentile) as a function of propagation time.}
\label{fig:error}
\end{figure}

\begin{figure}[H]
\centering
\includegraphics[width=\textwidth]{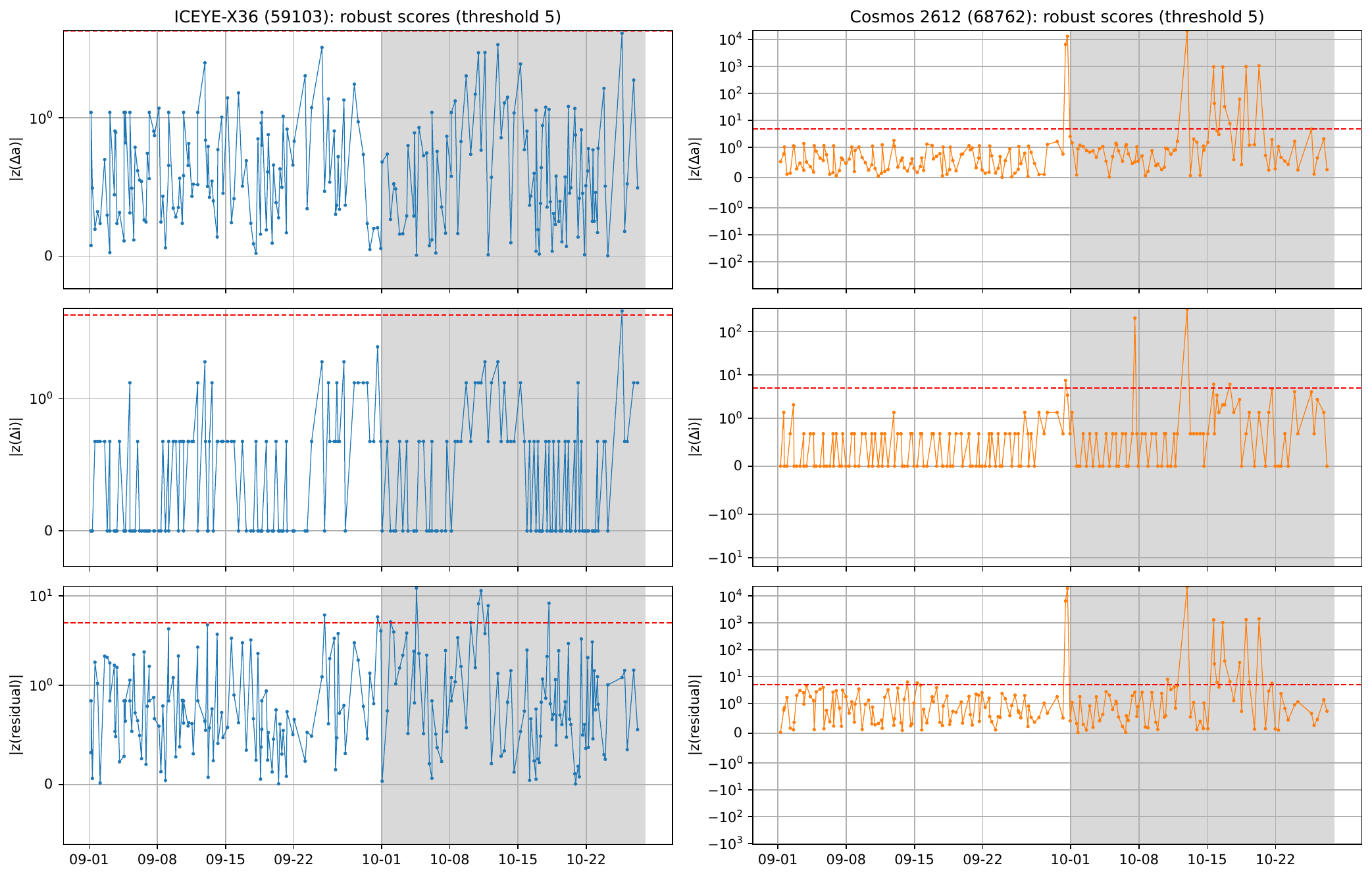}
\caption{Robust scores of the manoeuvre detector from 1 September for each object (threshold $|z| = 5$ in red).}
\label{fig:deteccion}
\end{figure}

\begin{figure}[H]
\centering
\includegraphics[width=0.8\textwidth]{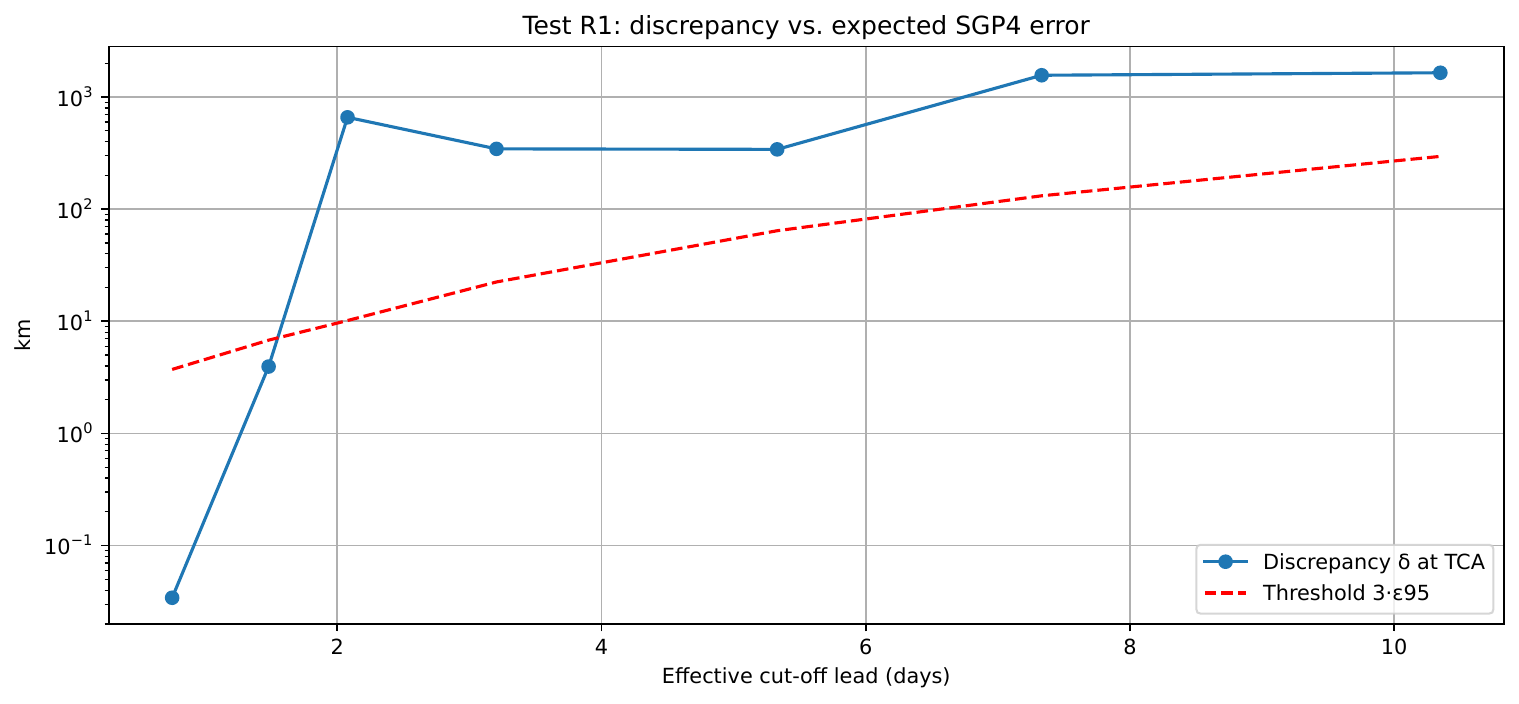}
\caption{Test R1: discrepancy at the TCA against the threshold $K\varepsilon_{95}$ as a function of the cut-off lead time.}
\label{fig:r1}
\end{figure}

\end{document}